# Many-Worlds Interpretations Can Not Imply 'Quantum Immortality'

Jacques Mallah, Ph.D.          (jackmallah@yahoo.com)     http://onqm.blogspot.com/

**Abstract:**

The fallacy that the many worlds interpretation (MWI) of quantum mechanics implies certain survival in quantum-Russian-roulette-like situations (the 'Quantum Suicide' (QS) thought experiment) has become common enough that it is now necessary to publicly debunk this belief despite the risk of further publicizing it. 'Quantum Immortality' (QI) is an extension of the QS Fallacy (QSF) with some additional unlikely assumptions. The QS/QI ideas are examined here and shown to be false.

**Introduction:**

For those familiar with the many worlds interpretation (MWI) of quantum mechanics, the 'Quantum Suicide' thought experiment is simple to describe, but in analyzing it there is a dangerous mistake that has often been made due to lack of understanding of the ways in which probabilities relate to the deterministic MWI. This mistake leads to the false conclusion that the MWI, in effect, implies certain survival.

A number of otherwise intelligent thinkers have fallen into this unfortunate pitfall, including Max Tegmark, who publicized the thought experiment in 1997. [Brooks] [Squires] may have been the first to publish a mention of it in his 1986 book. The 'Quantum Immortality' idea was believed by MWI's discoverer Everett long before that, possibly as early as 1957, although he did not publicize it. [Shikhovtsev]

Like Everett, many physicists have avoided writing about the idea, either because it sounds too crazy, or because they fear the real world consequences of risking spreading the idea even by publically criticizing it. I have been in the latter camp, but this 'Cat' is now clearly out of the bag, as an internet search will confirm. I have in the past participated in numerous internet discussions about the idea, but that was on specialized email discussion groups; the idea is now presented on high traffic websites such as Wikipedia.

This paper is meant to be the definitive debunking of the idea which can be cited as needed. I have no doubt that certain hard-core believers in 'Quantum Immortality' will never be swayed (and it would be akin to a religious conversion in more ways than one if they were), but this paper is not aimed at them. It is primarily aimed at those who find the MWI plausible and are ready to investigate its consequences.

**Quantum Russian Roulette:**

The QS thought experiment has been described as 'Schrödinger's Cat from the point of view of the cat' [Lewis], but the best known version is Tegmark's, which involves a 'quantum gun'. [Tegmark 1] This gun is used to play Russian Roulette, but instead of a

'classical' mechanism, a quantum measurement is used to determine whether the gun fires.

In a single-world interpretation, there is some known probability that the gun will fire, say 50%. If it does, the experimenter dies; if not, he survives unharmed.

In the Many Worlds Interpretation, both outcomes occur. There is no randomness involved; instead, there is 'branching' of the wavefunction. In this case, each of the two 'branches' has a total squared amplitude that is half as much of that of the original 'trunk' (which itself was some 'branch' of the universal wavefunction).

If the Born Rule holds for the MWI, an external observer would in effect assign a 50% chance that she would see the experimenter kill himself in the course of this experiment. At least superficially, this seems easy to understand: After the experiment, there is 'a copy' of the observer on each branch, and before the experiment for all practical purposes she can plan her life as though she is equally likely to end up as either of those copies. Also, if she closed her eyes and plugged her ears during the experiment itself, then just after the experiment, she is one of the copies but does not know which and must assign a 50% subjective chance of being either.

The experimenter himself is in a different situation. After the experiment, he will have 'a copy' on only one of the two 'branches' - the one in which the gun did not fire. It is at this point that the false conclusion – the Quantum Suicide Fallacy (QSF) - can seem to suggest itself: Should not he subjectively expect that he will always end up on the branch in which he survives? And in so expecting, then aside from any concern for the external observers (which is of course an important factor in practice), shouldn't he not mind conducting the experiment? And in finding that he survived multiple runs of the experiment, will not the experiment provide him (and only him) with further evidence for the MWI, as Tegmark claimed?

**Many Planets:**

In order to see why the QSF is false, it can be helpful to consider a different sort of many-worlds model, that of an infinite (or very large) classical universe. This has more similarity with the quantum mechanical case than one might think.

In a sufficiently large universe which is homogenous at large scales, there will be many Earth-like planets with creatures much like the humans of earth. Given a large enough sample of planets to work with, the resemblance can be made arbritrarily close. Thus, there will even be a large number of planets with individual people on them that are (aside from spatial location) identical to the individuals on Earth. These people (including the ones on Earth) will be called copies of those individuals.

The classical revolvers guns used in the Russian Roulette experiment by the copies of the experimenter will differ slightly, such that 50% of them will fire. Thus, 50% of these people will die with each run of the experiment.

For an individual 'copy of the experimenter', who is of course simply a human being in his own right, taking part in such an experiment is of course highly risky. Here, personal identity is tracked by following a physically distinct copy. Is this a significant difference between this case and the quantum mechanical case, since the experimenter's concern is not about the number of copies, but about the fate of the one copy that he personally is? Or is the more fundamental drawback to the QS experiment the fact that while many 'copies' that are much like him will survive, their population will be only 50% of what it was before?

Here is why the factor of physical distinctness is irrelevant: Just like with Star Trek's transporter beam, the experimenter would have no need to fear if he was made to vanish while another copy is simultaneously created elsewhere. More generally, assuming a physicalist/functionalist philosophy of mind (as MWIers generally do), personal identity over time is *not* a fundamental thing. Even the atoms that make up a person are not constant over time. And given the possibilities of personal fission and fusion, personal identity is not well-defined.

For most practical purposes an identity can be defined that depends on such things as causal chains and memories, but all copies have the same memories. Casual chains could be discussed in terms of modeling the individual as an implementation of an extended computation, but in the analogous quantum mechanical case, the number of such computations [Mallah 1] would correspondingly decrease in the 'quantum suicide' case. It is equally valid to eschew all definitions of personal identity extended over time; any conscious observation involves only a single observer-moment and that is the only fundamental thing.

Perhaps one might think that it does not then matter how many copies there are of an individual. However, if that were true, then for example a typical observer would not be likely to be a product of Darwinian evolution rather than spontaneous assembly. The latter class of observers is more varied than the former, but it certainly has many fewer copies of each type of individual in an infinite universe. (Or I should say a lower density of copies, if conceptually taking a ratio of infinite numbers upsets you.)

A more direct way to make the point is as follows: Suppose that the gunshots (the ones that go off) don't kill the unlucky copies of the experimenter immediately, but cause them to bleed to death over the course of several minutes. During that time, there are two classes of copies – 1) unharmed, and 2) dying. It should be obvious that in this case running the experiment was a bad idea.

After the dying is over, the surviving copies will in no way benefit from the deaths – consciousness can not magically jump into them from the dead copies. Thus the final state will also be an unfortunate one, since people have been killed. If the deaths are instantaneous, the final situation will not be any better than it was in the above case – the survivors are physically identical to those in the above case. Thus, what matters is that the total *amount of consciousness* has decreased.

If one denies that the amount of "a person's" consciousness *can* change as a function of time after it begins to exist and as long as there is at least *some* of it left, then in the quantum MWI, since there deterministically *is* some slight amplitude that any given particle configuration (such as that of a person's brain) exists even shortly after the Big Bang, there would again be no reason to expect that a typical person would be the result of normal evolutionary processes – you would have been 'born' way back then.

**Measure and Effective Probability:**

With regard to some observed outcome that can differ among people (including copies or near-copies of one person), an "effective probability" can be defined for classical mechanics in terms of the numbers of minds that see each outcome. Note: This mean that it is "in effect" a probability, playing the same role as a probability would have played for practical purposes such as making predictions, *not* that it is an actual probability which is producing some effect!

Explanations will be presented later in this paper as to ways in which the "effective probabilities" play roles similar to actual probabilities. However, it should be immediately clear that if there is a greater amount of consciousness seeing one outcome as compared to other outcomes, then that outcome must have a higher "probability" insofar as "probability" relates to the *commoness* of a given observation.

effective probability of an outcome = (# of minds that see that outcome) / (total #)

For MWI quantum mechanics, I have argued elsewhere that the number of implementations of conscious computations by the wavefunction plays the role of "number of minds" [Mallah 1] and is proportional to the squared amplitude in each branch, thus explaining the Born Rule in terms of observer counting. However, nothing here will hinge on that. What will matter is that the amount of consciousness in the different "branches" of the wavefunction differs.

In general, **the amount of consciousness**, analogous to the number of minds, is called the *measure* (of consciousness). It is conceivable that minds will not all have the same "amount of consciousness" (and this depends on one's theory of consciousness, details of which are not relevant here) in which case use the generalized formula

effective probability of an outcome = (measure of seeing that outcome) / (total measure)

When time variation is an issue, the relevant measure is the subjective-time integral of that, e.g. (the number of minds) x (the amount of subjective time they are conscious), which can be called the number of observer-moments. This latter will be called the *integral measure* if a distinction is needed for clarity. The distinction will not usually be made explicit. (Subjective time is used so that, for example, a fast conscious computer would have more integral measure per unit physical time than a slow one running the same algorithm would.)

Assuming the standard Born Rule holds, the measure in QM is proportional to the total squared amplitude of a branch (and to the number of classically distinct people who see the relevant observation within that branch). It is *not* proportional to the number of sub-branches.

To be precise, the Born Rule may be restated for use with the MWI in terms of measure as follows:

**Born Rule for a single classically distinct observer**:
The measure of seeing an outcome in MWI QM = (proportionality constant) x [Sum, over branches in which the observer sees that outcome, of the squared amplitude of such branches]

This can be generalized to the case of multiple classically distinct observers, which is useful for things like the anthropic principle (see below), as follows:

**Born Rule with possibly many classically distinct observers**:
The measure of seeing an outcome in MWI QM = (proportionality constant) x [Sum, over branches in which an observer sees that outcome, of {(squared amplitude of each such branch) x (# of classically distinct observers that see the outcome in that branch}]

**In an ordinary quantum mechanical situation** (**without births or deaths**) since the total squared amplitude is constant over time, **the total measure is constant over time**. In this case, the effective probability (per unit time) of seeing an outcome is proportional to the squared amplitude of that outcome (which is often taken as a statement of the Born Rule) and is independent of what happens in other outcomes.

For example, as a person grows older, and *assuming a safe environment* so there's no risk of death, there will be many 'branch points' in 'his life' (or more precisely, the various lives of 'his copies') and (assuming the MWI) 'he' will have 'split' into many (let's say N) 'different observers' with differing experiences. Does this mean 'his' total measure has increased N-fold? No, it doesn't; the squared amplitude 'he' started with originally is divided among the "N branches" and thus *the total measure summed over them is the same as 'he' started with*.

In situations such as 'Quantum Suicide' where death is involved, the effective probabilities (conditional on being after the experiment) are nonzero only in the branches in which the observer survived (i.e. number of people = 1). Since *total measure* (the denominator) *decreases*, the nonzero effective probabilities must be increased, in such a way as to normalize the total effective probability to 1 as per the definition above.

That is what can be misleading: because *usually* (when total measure is conserved) the effective probabilities are proportional to measure, it can seem natural to think about and use them as though there is no difference between effective probability and measure; but in general they are not the same thing and the difference is important.

The measures of consciousness are also nonzero only in the branches in which the observer survived, but they are **not** increased in those branches. *The total amount of consciousness **decreases** with each run of the QS experiment.* That is what matters.

**The Anthropic Principle:**

While the QS experiment could offer no experimental support for the MWI even to a surviving experimenter, there is a related class of observations which does legitimately provide support for some kind of multiple universes model. [Page] This is based on the Anthropic Principle, which notes that only universes that can support life will contain observers.

The difference between the two cases is as follows:

- In a single universe interpretation of QM, on average 50% of the people who perform the QS experiment will have survived. Their total measure is 50% of what it had been.

- In a MWI of QM, exactly 50% of the measure of people who perform the QS experiment will remain.

Suppose there are 10 billion people, and 200 of them decide to try QS, so about 100 of those survive. The effective probability of a person being any one of those QS survivors is about 100 in 10 billion. This is true in either the single-world or MWI case, so seeing that you are a QS survivor does not provide evidence either for or against the MWI.

Also in both cases, the effective probability of being alive (that is, seeing some observer-moment) prior to the deaths caused by the experiment is a bit higher than after, since there was more measure before.

But consider the observation that the universe seems fine-tuned to support life. Suppose that in a single universe case, the probability of this is some tiny number, X. The probability that any observers exist at all is X.

In a multiple universes case in which the physical laws or constants vary sufficiently, there will inevitably be some universes with just the right combination to support life. The fraction of such universes is X. If you are an observer, the effective probability that you will be in such a universe is 100%.

It is true that if you were the only observer that ever existed, and you tried many runs of the QS experiment, and survived, and lived for a long time afterward, that would provide evidence for the MWI similar to that of the Anthropic Principle case. In a single universe interpretation, your survival at all would have been unlikely. This is where the random aspect of the single-universe probability interpretation would make all the difference, because the "average" number of survivors would be less than one in the single-universe case, while in the MWI there are always numerous quantum branches.

However, even in such a case, you would not be justified in acting as though the QS Fallacy were true: Trying the experiment one more time would further reduce your measure by 50%. It would be just as though there were N people left before the experiment, and only N/2 afterwards.

**Effective Probabilities Deserve the Name:**

Decision theory has been used in attempts to derive the Born Rule for the MWI. I argue elsewhere that such attempts fail [Mallah 2], while the appearance of the Born Rule in the MWI might instead be explained or at least made viable by other means [Mallah 1]. However, it turns out that decision theory can still be useful for understanding the meaning and role of 'probabilities' in the deterministic MWI [Greaves]. I will assume here that for whatever reason the Born Rule *is* true for the MWI, in the sense that the measures of consciousness for each person are proportional to the squared amplitudes, and explain the consequences.

There are actually four complementary ways in which conditional "effective probabilities" (as defined above in terms of measure ratios) can play the role of 'probabilities' in the MWI. Each applies in a distinct context. Building on the work of Greaves, I will call these the Reflection Argument, Theory Confirmation, Causal Differentiation, and Caring Coefficients.

1) The Reflection Argument

The Reflection Argument applies when a measurement has already been performed, but the result has not yet been revealed to the experimenter. In this case he has *subjective uncertainty* as to which outcome occurred in the branch of the wavefunction that he is in, and he must assign some subjective probability to his expectations of seeing each outcome when the result is revealed.

He should set his subjective probabilities equal to the effective probabilities, because that way, the amount of consciousness seeing each outcome will be proportional to its subjective probability. If 2/3 of his copies (or measure) will see outcome A while the other 1/3 see B, he should assign a subjective probability to A of 2/3. If he does follow this policy consistently for many experiments, then most of his consciousnesses will be correct in the long run as to the average values of outcomes.

The Reflection Argument can be misleading when thinking about QS because it deals directly with probabilities after the experiment. It is important to remember that it applies only in a very specialized set of circumstances – after any splitting of observers caused by an experiment, but before the outcome is known. This does not apply when making decisions prior to an experiment because there is no subjective uncertainty in that case [Greaves].

2) Theory Confirmation

It may be than an experimental outcome is already known, but the person does not know what situation produced it. For example, suppose a spin is measured and the result is either "up" or "down". The probability of each outcome depends on the angle that the preparation apparatus is set to. There are two possible preparation angles; angle A gives a 90% effective probability for spin up, while angle B gives 10%. Bob knows that the result is "up", but he does not know the preparation angle.

In this case, he will probably guess that the preparation angle was A. In general, Bayesian updating should be used to relate his prior subjective probabilities for the preparation angle to take the measured outcome into account. For the conditional probability he should use for outcome "up" given angle A, he should use the effective probability of seeing "up" given angle A, and so on.

This procedure is justified on the basis that most observers (the greatest amount of conscious measure) who use it will get the right answer. Thus, if the preparation angle really was B, then only 10% of Bob's measure would experience the guess that A is more likely, and the other 90% will see a "down" result and correctly guess B is more likely.

Theory Confirmation can be applied in situations where time is a variable, using integral measure for effective probabilities; for example, if the observation in question is the time shown on a clock, and the hypotheses under consideration are whether the clock is accurate or inaccurate. This procedure of Theory Confirmation will be used below to argue that immortality is experimentally disproved.

3) Causal Differentiation

It may be the case that some copies of a person have the ability to affect particular future events such as the fate of particular copies of the future person. For example, in the Many Planets case, each copy's decision, together with external variables, determines the fate (such as life or death) of the future copy of him on the same planet.

It might seem at first that this would not apply to the MWI of QM, unless hidden variables are added to the picture. There are, in fact, hidden variable many-worlds models such as Continuum Bohmian Mechanics and as interpretations they have certain advantages of their own and may become more popular. However, there are ways that such a situation might occur even in the standard Everett ontology of just the wavefunction existing. For example, in the Many Computations Interpretation [Mallah 1], a mind may be associated with events taking place in a very particular part of configuration space and may not share the same fate as other minds on the same macroscopic "branch".

Pure Causal Differentiation situations are the most similar to classical single-world situations, since there is genuine ignorance about the future, and normal decision theory applies. Thus, if the man values the life of the future copy of him that he can affect, he should not risk suicide. QS would make no more sense than classical Russian Roulette.

It must be noted that no particular definition of personal identity is needed here. There is no fundamental difference between having the ability to affect the fate of one's future copy, as compared to having the ability to affect the fate of any other future person. That is why this is called Causal Differentiation, not Personal Differentiation.

4) Caring Coefficients

As opposed to Causal Differentiation, which may not apply to the standard MWI, the most standard way to think of what happens to a person when a "split" occurs is that of personal fission. In this case, since the MWI is deterministic, and assuming that the experimental setup (and in principle the wavefunction) and its physical consequences are known, decision theory that involves weighing the *future* consequences of a choice involves taking into account neither genuine randomness nor (subjective) uncertainty due to lack of knowledge.

Thus, the correct decision theory to use is simply to choose the outcome that has the most "utility" to the decider based on his emotional desires. (The term "utility" is unfortunate, since it suggests desirability *only* as a means to some other end. The term "desirability" would be more appropriate, but I will stick with the conventional term.)

It is important to note that one does not use some bastardized form of 'decision theory in the presence of subjective uncertainty' for this case. The decider may prefer to reason "as though" he will end up on one randomly selected branch of the wavefunction with appropriate probabilities, but in fact, he **can not** end up on just one – depending on definitions, he must either end up on all of them, or being just an observer-moment he will not exist in the future.

Rationality does not constrain utility functions, so at first glance it might seem that the decider's utility function might have little to do with the effective probabilities. However, as products of Darwinian evolution and members of the human species, many people have common features among their utility functions. The feature that is important here is that of "the most good for the most people". Typically, the decider will want his future 'copies' to be happy, and the more of them are happy the better.

In principle he may care about whether the copies all see the same thing or if they see different things, but in practice, most believers in the MWI would tend to adopt a utility function that is linear in the measures of each branch outcome:

$U_{total} = \Sigma_i \Sigma_p m_{ip}[\text{Choice}] q_{ip}$

where i labels the branch, p denotes the different people and other things in each branch, $m_{ip}$ is the measure of consciousness of person (or animal) p which sees outcome i, and is a function of the Choice that the decider will make, and $q_{ip}$ is the decider's utility per unit measure (quality-of-life factor) for that outcome for that person.

The measures and quality-of-life factors depend on time; the above expression for total utility should be integrated over time if the time variation is an issue. (For example, doubling the population of Earth might reduce the length of time the planet is inhabitable by a factor of four, as well as reducing the quality of life, and in that case would not be desirable.) The measures here can be called "caring measures" since the decider cares about the quality of life in each branch in proportion to them.

For cases in which measure is conserved over time this is equivalent to adopting a utility function which is linear in the effective probabilities. For cases like QS where measure is not conserved, it is not equivalent. How can we be sure that the correct utility function involves the measures directly, rather than the effective probabilities?

Admittedly, there is no such thing as a correct or incorrect utility function – it all depends on the decider's emotional desires. However, by asking questions such as whether a person would like to extend his life by reliving the exact same moments over again or would prefer to die, it can be determined whether an individual really would like to have a greater total amount of consciousness among his 'copies'. In all cases in which I have discussed the issue with QS advocates – and there were many – I have concluded that they really do want more measure and do not expect to lose measure with QS. The reason that they believe the QS fallacy is not that their utility functions are different, but only because they do not really understand the distinction between measure and probability.

The utility issue has been debated to some extent in the literature: [Papineau] wrote "If one outcome is valuable because it contains my future experiences, surely an alternative outcome which lacks those experiences is of lesser value, simply by comparison with the first outcome." In other words, he is saying that an outcome i where $m_{ip} = 0$ is of less utility than one with $m_{ip} > 0$. (Here, p = Papineau.) Thus, since QS would lead to more "$m_{ip} = 0$" outcomes (or reduce the measure of "$m_{ip} > 0$" outcomes), it is bad.

Papineau's analysis has been contested by [Tappenden] who wrote (and it must be noted that here the term "branch measures" means the squared amplitudes, not measures of consciousness as I have defined it) "… there is good reason to think that in quantum death-threat situations personal expectational weights part company with branch measures. The fact that utilities must track branch measures in non-death-threat situations does not provide a motivation for thinking that the same is true in death-threat situations …"

Tappenden's error is as follows. First, he invokes 'personal expectational weights', which are not directly relevent since there is no subjective uncertainty involved. To his mind this invocation lends force to his idea that utilities would not track "branch measures".

In precise terms, my interpretation of what he must have been thinking is that the measures of consciousness would not track the squared amplitudes of the remaining branches with the living observer in death-threat situations, but would instead be renormalized so that the total measure of consciousness remains constant as a function of time; that is, $\Sigma_i\ m_{ip} = M_p$ = constant. *If* that happened, it *would* justify his contention that utilities would not track squared amplitudes on the branches where life remains, since $m_{ip}$ on those living branches would suddenly increase when deaths occur on other branches (which may be when the dead branches are created).

However, there is no reason to believe that total measure of consciousness would be conserved in that way. From a functionalist philosophy of mind, each branch i is not physically influenced by events on other branches, so the amount of consciousness it gives rise to should not be affected by such events. The past history that led to the current branch also should not matter as long as the current physical characteristics of the branch are the same. Also, the equation $\Sigma_i\ m_{ip} = M_p$ = constant is poorly defined since persons are not well defined (how can one distinguish a diverged copy of you on a different branch from a guy who is merely similar to you?) The (implicit) notion that one's measure of consciousness must be conserved as a function of time is only an artifact of failure to distinguish measure of consciousness from probability.

**In summary, on effective probability and the QSF:**

There is *no* real randomness when QM splitting occurs in the MWI. *There is no consciousness that randomly follows either one track or the other, regardless of definitions.* Based on different possible definitions you can say *it does both*, **or** *it does neither*, and it *doesn't matter which one you say* because **real effects are definition-invariant**. You just can't use conflicting definitions at the same time.

Observations *that can be treated as if* "randomly chosen" for practical purposes come in to play for things like theory confirmation: as explained above, not because there's really something random involved whatsoever, but because doing so maximizes the number of observers who predict correctly and so is the most advisable thing to do.

By contrast to these facts, *to believe in the QS fallacy, you'd have to believe that 1) there is a consciousness that randomly follows one track* AND 2) which chooses only from tracks that continue in time. (Were these beliefs true, measure for the consciousness in question would always remain constant.)

**There is nothing like that on *either* count**. What confuses people is that despite that, for practical (decision-making) purposes only, you can think of point 1) as giving the correct answers, as long as you reject point 2).

QSF believers are liable to say that "from a first person point of view" it seems as if there is a consciousness that randomly follows one track, chosen from the continuing ones, and claim that this is enough to justify the QS idea. But that claim is nonsense, as any of the ways in which randomness "seems" to come into play are fully explained as properties of measure-ratio-based effective probabilities as explained above, and clearly there is **no** *actual* random following of one track (chosen from continuing tracks only, and thus preserving its total measure) as would have been required to justify the QS idea.

**'Quantum Immortality':**

The "Quantum Immortality" Fallacy (QIF) is an extension of the QS fallacy to include all forms of death. Since there is always some small probability that any dangerous situation will be survived, there will in general be some small amplitude branch of the wavefunction in which that is the case, so in a sense all threats are quantum threats. Most advocates of "quantum immortality" go further and argue that the existence of any mind (at any time) which is like a subjectively future version of one's current mind, even in the absence of any casual link, is enough to guarantee personal escape from death. This ignores the lack of definition of personal identity related to similar minds.

Mind B, that could be a future version of another mind A, can seem to be a continuation of A as far B can tell. It could be said in some sense that "from a first-person point of view" it seems like the consciousness that was associated with mind A is now associated with mind B. Thus, there is the illusion of a flow of consciousness from A to B. The QIF advocates consider such a flow reified (or at least they do so in practice, though they may deny it), to the extent that (in effect) they believe that the total measure associated with minds like B must be equal to that associated with potential previous minds like A.

For example, according to this view, if in the distant past there was one copy of Bob, and then on a million distant planets new copies of Bob come into being through normal evolutionary processes, the Bobs' total measure nonetheless remains constant and is presumably equal to that of just one person rather than that of a million people. (QS is merely the same kind of process in reverse.) It is not at all clear why, in the QIF view, a single distant precursor leads to such a low measure for each current Bob, while presumably having zero distant precursors would not have lead to low or zero measure for the Bobs.

All of the arguments against the QS fallacy apply against the QIF. In addition, there are two new arguments to be made against it: A general argument that people are not immortal, and Tegmark's view on gradual decrease of quality of consciousness.

**A General Argument Against Immortality:**

The method of Theory Confirmation can be applied to the question of immortality. In general, if we are immortal, there would be two classes of observations: Those made by normal people within a normal lifespan, and those made in the 'afterlife'. For 'quantum immortality' the 'afterlife' will be taken to mean those who find themselves to be much older than a normal human lifespan.

If the 'afterlife' is infinite, then it will have infinitely more integral measure than the normal life. Thus, the effective probability of finding oneself in a normal lifetime would be zero. If there is no 'afterlife' then the effective probability of that would be unity. By applying Bayesian reasoning, this implies that if one *does* find oneself in a normal lifetime, as we do, there must be no infinite afterlife.

Note that this does not mean that there can't be an infinitely long (in time) tail in which measure decays, just that the ratio of the integral measure in the tail to the integral measure in a normal lifetime must be finite, such as with an exponential decay of the survival fraction. Since the effective probability of being older than a given age approaches zero for older ages, this certainly is not immortality in any meaningful sense.

It is sometimes objected that even if there is an infinite afterlife, the normal lifetime will still exist, and in that case those within the normal life who apply the reasoning would be wrong. Indeed, while the reasoning is certain to work for almost all observations, it will fail for this set. The set has effective probability zero, but still exists, just as the chance of randomly choosing a real number from 0 to 10 with a uniform distribution and getting $\pi$ is zero but that does not mean that $\pi$ is any less likely than any other number in the range, and in an MWI, it will happen in some set of worlds. The procedure of using effective probabilities for Theory Confirmation will not work correctly for all observations, but with a zero chance of failure in this case, it is still far more reliable than any of our theories of physics.

If there is an 'afterlife' of long but finite duration, similar reasoning applies, but in that case the effective probability of being in a normal lifetime is merely small, not zero. While not conclusive, this would still be strong evidence against that possibility.

This proof still leaves the possibility that the MWI is false and that we would have been immortal if only it had been true. However, advocates of "quantum immortality" believe the MWI and QI are true.

**Max Tegmark's views:**

Max Tegmark publicized the QS idea, but in some ways he is more of a moderate on the issue than most of its believers are. If he were to follow in the footsteps of Don Page and alter his views, recanting belief in QS, it would be a great help in exposing the belief as a fallacy, and I hold out hope that it is possible that he will do so.

In his paper [Tegmark 1] QS is explained as follows:

$$U \frac{1}{\sqrt{2}} \left( |\uparrow\rangle + |\downarrow\rangle \right) \otimes |\ddot{\ }\rangle = \frac{1}{\sqrt{2}} \left( |\uparrow\rangle \otimes |\ddot\smile\rangle + |\downarrow\rangle \otimes |\overset{\times\times}{\smile}\rangle \right)$$

"Since there is exactly one observer having perceptions both before and after the trigger event, and since it occurred too fast to notice, the MWI prediction is that" (the experimenter) "will hear "click" with 100% certainty."

That is a rather odd statement because he is certainly aware that in the MWI there is no sense in which it can be rightfully said that "there is exactly one observer" either before or after the experiment. The ket notation may be unhelpful here; indeed, if the tensor product of kets on the left hand side were expanded instead of factoring out the observer, there would appear to have been "two observers" initially.

In fact, there are countless sub-branches to each "world" and one might think that each of these supports one observer, but even that does not tell the whole story, because if it did the Born Rule would not hold. The only promising form of observer-counting for the MWI that I am aware of is the Many Computations Interpretation (MCI) [Mallah 1], and in that context, if the Born Rule holds normally, it may be said that the number of observers (computations) after the QS experiment is *half* of what it was before.

More generally, whether or not observer-counting per se can be performed, what is true is that the measure (amount) of consciousness after the experiment is half of what it was before. For all practical purposes, this is equivalent to a 50% chance of death.

Otherwise, if the "number of observers" in the |up> branch alone *is* considered equal to the original total, then in an ordinary (non-deadly) spin measurement, yielding observers in the |down> branch *in addition to* the |up> ones, the number of observers would double. In other words, measuring a spin normally would be equivalent to cloning yourself (in the sci-fi sense). This would have two notable consequences: 1) Your measure would dramatically increase as time goes on, so it is very unlikely that you would not observe yourself to be very old, and 2) measuring spins would be equivalent to having babies in terms of creating human life, and failure to do so would be criminal in terms of opportunity cost, especially if you are in fact creating whole new universes!

While Tegmark believes the QS fallacy and believes the MWI is true, he does not believe in "quantum immortality". He wrote the following explanation [Tegmark 2]:

*"Here's a brief comment on the issue of whether the MWI implies subjective immortality. This has bothered me for a long time, and a number of people have emailed me about it after the Guardian and New Scientist articles came out. I agree that if the argument were flawless, I should expect to be the oldest guy on the planet, severely discrediting the Everett hypothesis. However, I think there's a flaw. After all, dying isn't a binary thing where you're either dead or alive - rather, there's a whole continuum of states of progressively decreasing self-awareness. What makes the quantum suicide work is that you force an abrupt transition. I suspect that when I get old, my brain cells will gradually give out (indeed, that's already started happening...) so that I keep feeling self-aware, but less and less so, the final "death" being quite anti-climactic, sort of like when an amoeba croaks. Do you buy this?"*

Tegamark goes on to explain that in his view QS would only work if it's a genuinely quantum event (ruling out regular-gun Russian roulette) and if death occurs before the person in the doomed branch can become aware that he is doomed.

I was encouraged to see that Tegmark realized that he "should expect to be the oldest guy on the planet" if QI were true; in context, that is an elegant but approximate version of the general argument against immortality given above. My response to him was this:

"Max … Why can't you realize that even if the amoeba … disappeared completely, or if you were to consider the amoeba to be some individual other than yourself, it would not help you in the slightest to be effectively immortal?"

Meaning: Removing something from one branch of the wavefunction has no effect on the other branches. **The "ameoba" in no way diminishes the consciousness in the other branches that have fully conscious copies of the person, nor would they be any different if the ameoba branch had been an abruptly dead branch instead.** The reason that he would not be immortal is simply that the fully conscious branches, in and of themselves, have measure that diminishes over time.

**'QS' ideology is dangerous:**

As revealed in the internet discussions, QS is not just an academic issue. People have considered putting the experiment into practice, and have done such things as carry a real gun into a casino while so considering; perhaps some have gone all the way. (Liz Everett's suicide may have been a variant of it. [Shikhovtsev]) The reasons that most (if not all so far) people who consider QS have backed off are concern for those left behind, fear of botched suicide attempts, doubts about the MWI, and doubts about QS within the MWI – in that order.

What is worse, is that while the discussion has generally been framed in terms of suicide, there is nothing to prevent its believers from proceeding to "quantum murder": e.g. If a man is very poor, then they may kill him, in the belief that "he may end up in" a better place. [Higgo 2] Thus, the QS/QI fallacies can threaten to lead to nothing less than a sort of postmodern fanatical religious cult, complete with promises of immortality, suicides (perhaps by willing suicide bombers), and murder.

For a philosopher of physics interested in the MWI of QM, this situation poses an unfortunate dilemma: Promoting belief in the MWI will also increase the number of people who believe the QS fallacy, and ultimately will cause deaths, unless the fallacy is debunked with sufficient vigor and persuasiveness. The present situation, in which it has become common for people who study quantum interpretations to believe that the MWI would imply immortality, can not be tolerated.

The QS fallacy resembles other dangerous pseudoscientific fallacies such HIV/AIDS denialism and the use of worthless homeopathic "medicine" to treat serious illness. In these other cases, a hard core of believers has always remained in spite of all evidence and scientific logic. There is no reason to doubt that the same would apply here. Even in the future, when (I am sure) the QS fallacy will be more widely seen for what it is, real scientists will have to continue to speak out against this dangerous pseudoscience in the strongest terms.